\documentclass{sf2a-conf2010}
\usepackage{graphicx}
\usepackage{hyperref}
\usepackage[]{natbib}

\begin{document}
\TitreGlobal{SF2A 2010}
%
\title{MAGNETIC FIELD DRAGGING IN ACCRETION DISCS}
\author{R. de GUIRAN}
\address{Laboratoire d’Astrophysique de Grenoble, UMR 5571 CNRS, Universite  Joseph Fourier, BP 53, 38041 Grenoble, France}
\author{J. FERREIRA$^1$ }

\runningtitle{Magnetic field dragging in accretion discs }
%
\setcounter{page}{237}
\index{de GUIRAN, R.}
\index{FERREIRA, J.}

\maketitle
\begin{abstract}
%
%
Accretion discs are composed of ionized gas in motion around a central object. Sometimes, the disc is the source of powerful bipolar jets along its rotation axis. Theoretical models invoke the existence of a bipolar magnetic field crossing the disc and require two conditions to produce powerful jets: field lines need to be bent enough at the disc surface and the magnetic field needs to be close to equipartition. The work of Petrucci et al (2008) on the variability of X-ray binaries supposes that transitions between pure accretion phases and accretion-ejection phases are due to some variations of the disc magnetization. This rises the problem of the magnetic field dragging in accretion discs. We revisit the method developed by Lubow et al (1994) by including momentum and mass conservation equations in a time-dependent 1D MHD code. 
\end{abstract}
\section{Introduction}
  First works on magnetic field dragging in accretion discs have focused on the \citet{Blandford_&_Payne} necessary condition for cold jet launching, namely:
$tan^{-1}(B_r/B_z)\geq 30° $. Thus, \citet{1989ASSL..156...99V} , \citet{Lubow_&_al}, \citet{Reyes_&_al} developed different methods to address this question in a standard accretion disc (here after SAD). They all conclude that a large effective Magnetic Prandtl number  (here after $Pm=\nu_v/\nu_m$ as the ratio of the turbulent viscosity to the magnetic diffusivity) is necessary to reach a sufficient magnetic flux accumulation that satisfy the \citet{Blandford_&_Payne} criterion. Indeed, in a SAD, bent field lines requires a magnetic Reynolds number $R_m=rU_r/\nu_m=R_e P_m \sim r/h $ (with $R_e$ the Reynolds number), whereas in a SAD, $Re \sim 1$. This is problematic as it tends not to be satisfied according to 3D MRI simulations \citep{Lesur&Longaretti2009}. More recent works, by \citet{Lovelace2009}, showed that significant advection of magnetic field could be possible even with $Pm \sim 1$.\\
On the other hand, the existence of jets do require the presence of a large scale vertical field \citep{McKinney&Blandford}. It has been shown that this field must be close to equipartition in order to drive powerful jets \citep{Ferreira&Pelletier95}.\\
In this work, we revisit the standing problem of field advection in accretion discs, by introducing new physical input from modern works.

\section{Analytical description}
The full MHD equations could be written in axisymetry in the in the $(r,\phi,z )$  cylindrical coordinates.
Writting: $ a=rA_{\phi} $ with $\vec{\nabla} \times \vec{A}=\vec{B} $ and $b=rB_{\phi}$, we have:

\begin{eqnarray}
&& \frac{\partial \rho}{\partial t} + \nabla \cdot \rho \vec u = 0 \label{eq:mass}  \\
&& \frac{\partial \rho \Omega r^2 }{\partial t} + \nabla \cdot \left ( \rho \vec u_p\Omega r^2 - \frac{b}{\mu_o} \vec B_p \right) =  \frac{1}{r} \frac{\partial}{\partial r} \rho \nu_v r^3 \frac{\partial \Omega}{\partial r} \label{eq:mhd} \\
&& \frac{\partial a}{\partial t} + \vec u_p \cdot \nabla a = \eta_m J_{\phi} \label{eq:flux}
\end{eqnarray}
Where $\eta_m= \mu_0 \nu_m$\\
\\
We focus on the magnetic flux transport coupled with mass and momentum conservation. We assume a thin keplerian disc with $\varepsilon=h/r$ constant with radius ($h$ the height of the disc at a given radius). We also assume that only the turbulent viscosity plays a role in the angular momentum transfer.

\subsection{Induction equation}
We start with the induction equation, by setting $\Psi = \frac{1}{2h} \int a dz$, we rewrite equation \ref{eq:flux} integrated over the thickness of the disc:
\begin{equation}\label{mag-flux}
\frac{\partial \Psi}{\partial t} =  U_0\frac{\partial \Psi}{\partial r} - \frac{\eta_m}{2\varepsilon} J_{\phi S}
\end{equation}
With $U_0=\dot M_a/(2\pi r \Sigma)$ with $\Sigma$ the surface density, $\dot M_a$ the accretion rate and $J_{\phi S}$ the surface current density.
All the question is determining $J_{\phi S} $ in term of the radial distribution of $\Psi$. Approximating the disc as rings of toroidal current, an equivalence between $\Psi(r)$ and $J_{\phi S} $ can be found when assuming a potential magnetosphere \citep{Lubow_&_al}.

\subsection{Momentum conservation}
By considering a quasi Keplerian motion, one can deduce the accretion rate from equation \ref{eq:mhd}

\begin{eqnarray}\label{Mdot}
\dot M_a =\frac{6 \pi}{\Omega_K r} \frac{\partial }{\partial r} \left ( \Sigma \nu_v \Omega_K r^2 \right )
\end{eqnarray}

\subsection{Mass conservation}
We then use the mass conservation (equation \ref{eq:mass}) integrated over the thickness of the disc:
\begin{equation} \label{cons}
\frac{\partial \Sigma}{\partial t} = \frac{1}{2\pi r}  \frac{\partial \dot M_a}{\partial r}
\end{equation}

\subsection{Turbulent prescription}
At last, we need to do an assumption on the transports parameters: 
\begin{eqnarray}
\alpha_v &\equiv & \frac{\nu_v}{\Omega_K h^2} \\
{\cal P}_m &\equiv& \frac{\nu_v}{\nu_m}
\end{eqnarray}
Magnetic fields in accretion discs are known as a potential source of  turbulence via the Magneto-Rotational Instability (here after MRI). This is expected to lead to anomalous transport coefficients $\nu_v$ and $\nu_m$.  
Recent results by \citet{Lesur&Longaretti2009} show that $P_m \sim 2 $ and a behaviour of alpha in $\mu^{0.5}$ for low $\mu$ (with $\mu$  the magnetization: ratio of the magnetic pressure on  the thermal pressure). As we know that MRI is suppressed around equipartition,
we use the prescription:

\begin{equation}\label{alpha}
\alpha_v(\mu)=2(\mu(1-\mu))^{1/2}
\end{equation}
Where the local disc magnetization is computed as:
\begin{equation}
\mu = \frac{B_z^2}{\mu_o P} =  \frac{2 B_z^2}{\mu_o \varepsilon \Sigma \Omega_K^2 r} 
\end{equation}
One can study the evolution of the magnetic field and the density in the disc, by solving equations \ref{mag-flux}, \ref{cons}, using \ref{Mdot} and the equivalence between $J_{\phi S}$ and $\Psi(r)$.

\section{Results}
\begin{figure}[ht!]
 \centering
\includegraphics[scale=0.8,width=0.8\textwidth]{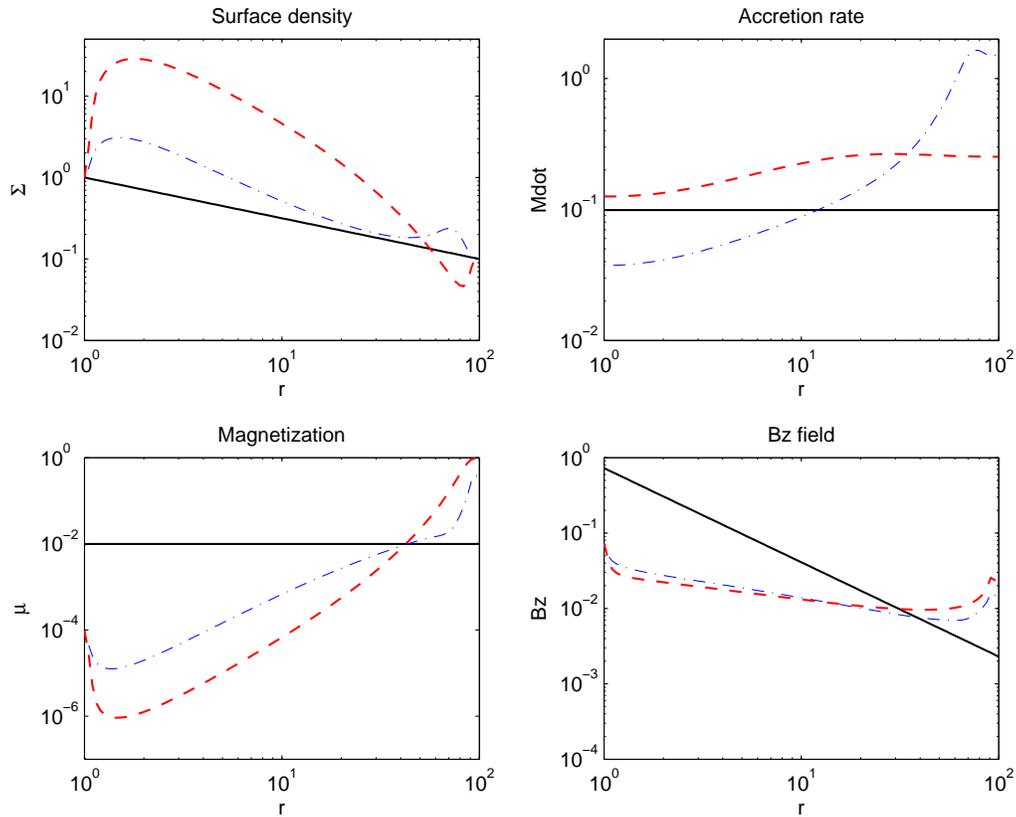}  
  \caption{Density, accretion rate, magnetization and magnetic field in the disc at 3 different times. All the quantities are normalized. The solid line corresponds to the initial state, the dash-doted line corresponds to 10 keplerian times at the outer radius, and the dashed line to 100 keplerian times at the outer radius.}\label{DEGUIRAN:fig1}
\end{figure}

We study the behaviour of a SAD with $P_m=1$ initially thread by a magnetic flux very compressed in the inner zone of the disc. This field is chosen to have a uniform magnetization at $t=0$. This  initial condition could model the high soft state (labelled C in \citet{2008MNRAS.385L..88P}), right after the jet quenching of X-ray binaries.
We use the $\alpha_v(\mu)$ prescription (equation \ref{alpha}). The results are shown on figure \ref{DEGUIRAN:fig1}. At the beginning, of the simulation, the magnetic field diffuse outward as expected in a SAD. Leading
to an increasing of $\mu$ in the outer zone of the disc, and so a noticeable positive variation of the accretion rate at
the outer edge of the disc. We can notice that at the end of the simulation, the disc is approaching a stationary state with an
accretion rate quite uniform at the outer radii of the disc. But as the magnetization is very weak in the inner zone, so is $\alpha_v$,
and then the dynamics are very slow.\\
One can also see that a strong magnetization is reached in the outer disc ($\mu \sim 1 $). This is very interesting because the condition
for a powerful jet launching is satisfied at these radii.Then, the dynamic of the disc becomes considerably different (which is not taken into account
in the code). If we consider the magnetic torque induced by the jet, we would assist to a new advection of the field from the outer edge of the disc.

\section{Conclusions}
The new advection of magnetic field supposed to occur at the end of the simulation would make a transition with the diffusion of the magnetic field observed in the simulation. Thus, we could assist to a phenomenon of magnetic tide. In this sense, this result is very encouraging for the study of transition from high soft to low hard state in X-ray binaries. However here, the accretion rate remains roughly constant during the simulation. But if we work by imposing the accretion rate at the outer radius, a decreasing of it would enhance the magnetization at the outer radii, which is compatible with a transition between high soft to low hard state.

\begin{acknowledgements}
Thank you to all the team from the laboratoire d'astrophysique de Marseille for the organisation of this conference.
\end{acknowledgements}

%
%
%
%
%

\bibliographystyle{plainnat}
\bibliography{biblio}

\end{document}